\documentclass[10pt]{article}
\usepackage{geometry}                
\usepackage[parfill]{parskip}
\usepackage{amssymb}

\usepackage{amsmath}

\title{\bf Equivalent power law potentials}

\author{C. V. Sukumar \\{\em Wadham College,}\\{\em University of Oxford, Oxford OX1 3PN, U.K. }
}

\begin{document}
\maketitle

\begin{abstract}
It is shown that the radial Schr{\"{o}}dinger equation for a power law potential and a particular angular momentum may be transformed using a change of variable into another Schr{\"{o}}dinger equation for a different power law potential and a different angular momentum. It is shown that this leads to a mapping of the spectra of the two related power law potentials. It is shown that a similar correspondence between the classical orbits in the two related power law potentials exists. The well known correspondence of the Coulomb and oscillator spectra is a special case of a more general correspondence between power law potentials. 

\end{abstract}

\section{Introduction}

In this study we investigate the circumstances under which Classical dynamics and Quantum Mechanics induce relationships between classical orbits, eigenvalue spectra and phaseshifts of different dynamical systems. We study in particular power law potentials which belong to a special category in the sense that they permit a unique scaling of length and energy which lead to dimensionless equations in Classical and Quantum Mechanics. The dimensionless equations allow certain transformations which reveal a connection between the orbits and spectra of two related power law potentials. It has been noted in earlier literature that the solutions to the Schr{\"{o}}dinger equation for a Simple Harmonic Oscillator (SHO) may be related to the boundstate solutions in a Coulomb potential, a relationship that is also evident in the connection between Hermite polynomials and Laguerre polynomials in the mathematical literature. In this report we examine the possibility of more general connections between power law potentials. In section 2 of this paper we study the relation between the Schr{\"{o}}dinger equations of two related power law potentials. This issue was discussed by  Quigg and Rossner in Physics Reports {\bf 56} (1979) in their study of quarkonium states using non-relativistic Quantum Mechanics. We follow their method of analysis and extract additional features which could prove useful.  In sections 3 and 4 we study the same issue from the point of view of semi-classical and Classical Mechanics and attempt to establish which of the features that appear in Classical Mechanics are preserved in the passage to Quantum Mechanics.

\section{Power law potentials in Quantum Mechanics}

We start from the radial Schr{\"{o}}dinger equation for a power law potential with power exponent $\nu_1$ and angular momentum $l_1$
\begin{equation}
\frac{\hbar^2}{2\mu_1} \ \frac{\partial^2u_1}{\partial r^2}\ +\ \Big(E_1\ -\ \lambda_1\ r^\nu\ -\ \frac{l_1(l_1+1}{2\mu_1 r^2}\ \Big) \ u_1\ =\ 0 
\end{equation}
For power law potentials a scaling length and a scaled energy may be identified 
\begin{equation}
a_1\ =\ \Big(\frac{\hbar^2}{2\mu_1|\lambda_1|}\Big)^{\frac{1}{\nu_1+2}}\ ,\ r\ =\ a_1\ \rho_1\ ,\ E_1\ =\ \frac{\hbar^2}{2\mu_1 a_1^2}\ \epsilon_1
\end{equation}
in terms of which the Schr{\"{o}}dinger equation in dimensionless form becomes
\begin{equation}
\frac{\partial^2 u_1}{\partial\rho_1^2}\ +\ \Big(\epsilon_1\ -\ \rho_1^{\nu_1}\ -\ \frac{l_1(l_1+1)}{\rho_1^2} \Big)\ u_1\ =\ 0 \label{eq:I0}
\end{equation}
We now introduce a new variable and a new function through the relations
\begin{equation}
\rho_1^{\nu_1}\ =\ z^{-\nu_2}\ ,\ u_1(\rho_1)\ =\ z^{-\frac{\nu_1+\nu_2}{2\nu_1}}\ v(z)\label{eq:I1}
\end{equation}
to transform the radial equation to the form
\begin{equation}
\Big(\ \Big(\frac{\nu_1}{\nu_2}\Big)^2 z^{2\big(1+\frac{\nu_1}{\nu_2}\big)} \Big[\frac{\partial^2v}{\partial z^2} + \Big(1 - \frac{\nu_2^2}{\nu_1^2}\Big) \frac{v}{4z^2}\Big] + \Big[\epsilon_1 - z^{-\nu_2} - \frac{l_1(l_1+1)}{z^2} z^{2\big(1+\frac{\nu_1}{\nu_2}\big)}\Big] v\Big)\ z^{-\frac{\nu_1+\nu_2}{2\nu_1}} \ =\ 0 
\end{equation}
If we now impose the conditions
\begin{align}
&2\Big(1 +\ \frac{\nu_2}{\nu_1}\Big)\ +\ \nu_2\ =\ 0\ \rightarrow\  \frac{1}{\nu_1}\ +\ \frac{1}{\nu_2}\ +\ \frac{1}{2}\ =\ 0 \label{eq:I2}\\
&l_1(l_1+1) + \frac{1}{4}\Big(1 - \frac{\nu_1^2}{\nu_2^2}\Big) =  l_2( l_2+1)\ \frac{\nu_1^2}{\nu_2^2}\ \rightarrow\ \big(l_1 + \frac {1}{2}\Big)^2 \nu_2^2\ =\ \Big(l_2 + \frac{1}{2}\Big)^2 \nu_1^2 \label{eq:I3}
\end{align}
the resulting equation is
\begin{equation}
\frac{\partial^2v}{\partial z^2}\ +\ \Big(-\frac{\nu_2^2}{\nu_1^2}\ +\ \epsilon_1 \frac{\nu_2^2}{\nu_1^2}\ z^{\nu_2}\ -\ \frac{l_2(l_2+1)}{\rho_1^2}\Big) v\ =\ 0
\end{equation}
A new scaling length and scaled energy defined by
\begin{equation}
v(z) = u_2( \rho_2)\ ,\ z = a_2\ \rho_2\ \ ,\ a_2^{\nu_2+2}\ \epsilon_1 \Big(\frac{\nu_2}{\nu_1}\Big)^2\ =\ 1\ ,\  \epsilon_2 = - a_2^2\ \frac{\nu_2^2}{\nu_1^2}\ =\ - \epsilon_1^{\frac{\nu_1}{\nu_2}}\Big(-\frac{\nu_1}{\nu_2}\Big)^{\nu_1} \label{eq:I4}
\end{equation}
may now be identified which leads to the transformed radial equation 
\begin{equation}
\frac{\partial^2 u_2}{\partial \rho_2^2}\ +\ \Big( \epsilon_2\ +\ \rho_2^{\nu_2}\ -\ \frac{l_2(l_2+1)}{\rho_2^2} \Big)\ u_2\ =\ 0  \label{eq:I5}
\end{equation}
 
Comparison of (\ref{eq:I0}) and (\ref{eq:I5}) shows that the radial equation for a confining potential with $0\le \nu_1 \le \infty$ and $\epsilon_1$ positive can be transformed to a new radial equation for an attractive singular potential with $-2 \le \nu_2\le 0$ , $ \epsilon_2$ negative and the eigenvalue spectra of the confining potential and the singular potential are related  by (\ref{eq:I4}). 

If the exponents are restricted to lie in the range $[-2,\infty]$ (\ref{eq:I2}) guarantees that $\nu_1$ and $\nu_2$ are always opposite in sign  and the choice
\begin{equation}
\Big(l_1 + \frac {1}{2}\Big)\ \nu_2\ =\ -\Big(l_2 + \frac{1}{2}\Big)\ \nu_1 \label{eq:I6}
\end{equation}
which is consistent with (\ref{eq:I3}) guarantees that a positive value of $l_1$ leads to a positive value of $l_2$ and ensures that the solution of the transformed equation vanishes as $\rho_2\rightarrow 0$. Upto this point our derivation parallels the derivation given by Quigg and Rossner in Physics Reports {\bf 56} 1979, pages 191-2. It is possible to take this discussion further which we now proceed to do.

If in addition (\ref{eq:I6}) transforms an integer value of $l_1$ to an integer value of $l_2$ then the mapping we have discussed relates the eigenvalue spectrum of a potential representing a possible physical system to the eigenvalue spectrum of another potential representing another possible physical system. If $l_1$ and $l_2$ are positive integers with $l_1 > l_2$ then the spectrum of a confining power law potential with positive exponent $\nu_1$ and angular momentum $l_1$ is related to the spectrum of an attractive potential with negative exponent $\nu_2$ and angular momentum $l_2$ if the exponents are such that
\begin{equation}
\nu_1 = \frac{4(l_1-l_2)}{2l_2+1}\ ,\ \nu_2 = -\frac{4(l_1-l_2)}{2l_1+1}\ ,\ \rightarrow\  \epsilon_2\ =\ -\Big(\frac{2l_1+1}{2l_2+1}\Big)^{\nu_1}\ \Big(\frac{1}{\epsilon_1}\Big)^{\frac{2l_1+1}{2l_2+1}} \label{eq:I7}
\end{equation}

A simple example of this relationship is realized by the choice $l_1=3l_2+1$ which yields $\nu_1=4,\ \nu_2 =-\frac{4}{3}$ and gives rise to the identification that the  eigenvalues for $l_1=1,4,7,..$ in the potential $\rho_1^4$ are related to the eigenvalues for $l_2=0,1,2,..$ of the attractive potential $-(\rho_2)^{-\frac{4}{3}}$ by the mapping $\epsilon_2 = - \frac{3^4}{{\epsilon_1}^3}$.

Another example of this relationship is realized by the choice $l_1=5l_2+2$ which yields  $\nu_1=8,\  \nu_2 =-\frac{8}{5}$ and leads to the identification that the eigenvalues for $l_1=2,7,12,..$ in the potential $\rho_1^8$ are related to the eigenvalues for $l_2=0,1,2,..$ of the attractive potential $-(\rho_2)^{-\frac{8}{5}}$ by the mapping $ \epsilon_2 = -\frac{5^8}{{\epsilon_1}^5}$. 

The two examples we have given illustrate the equivalence of two sets of power law potentials but it is evident that (\ref{eq:I7}) provides an entire family of pairs of potentials whose eigenvalue spectra are related. We have shown that the power law potentials $V_1=\rho_1^{\nu_1}$, $0\le\nu_1\le\infty$ and $V_2=-\rho_2^{\nu_2}$, $-2\le\nu_2\le0$ have related eigenvalue spectra if the conditions
\begin{equation}
(\nu_1 + 2)\ (\nu_2 + 2)\  =\ 4\ \ {\hbox {and}}\ \ \frac{2l_1+1}{\nu_1}\ +\   \frac{2l_2+1}{\nu_2}\ =\ 0 \label{eq:G8}
\end{equation}
are satisfied and the spectral relation may be given in the form
\begin{equation}
{\sqrt{\nu_1+2}}\ \big(\epsilon_1\big)^{\frac{1}{\nu_2}}\ =\ {\sqrt{\nu_2+2}}\ \big(-\epsilon_2\big)^{\frac{1}{\nu_1}} \label{eq:I8}
\end{equation}
which may also be given in the form
\begin{equation}
\nu_1\ \log|\epsilon_1|\ -\ \Big(\frac{\nu_1^2}{\nu_1+2}\Big)\ \log(\nu_1+2)\ =\ \nu_2\ \log|\epsilon_2|\ -\ \Big(\frac{\nu_2^2}{\nu_2+2}\Big)\ \log(\nu_2+2)
\end{equation}
exhibiting a symmetrical structure of the mapping.

\subsection{Relation between the spectra of the Oscillator and the Coulomb potentials}

The choice $\nu_1=2$ corresponds to the radial equation for the oscillator potential and it may be shown that for any values of angular momenta $l_1$, whether integer or not, the solutions of (\ref{eq:I0}) may be given as
\begin{equation}
u_1\ =\ \rho_1^{l_1+1}\ \Big(\exp \frac{-\rho_1^2}{2}\Big)\ M\Big(\frac{2l_1+3-\epsilon_1}{4},l_1+\frac{3}{2},\rho_1^2\Big) \label{eq:I9}
\end{equation}
where $M(a,b,z)$ is one of the solutions of Kummer's equation (Abramowitz and Stegun 1965) in the form
\begin{equation}
M(a,b,z)\ =\ 1\ +\ \frac{a}{b} z\ +\ \frac{a(a+1)}{b(b+1)} z^2\ +\ ...
\end{equation}
with polynomial structure when $a=-n$, where $n$ is an integer $\ge 0$. Hence for energies $\epsilon_1=4n+2l_1+3$ the solution for $u_1$ given in (\ref{eq:I9}) is normalizable and it satisfies boundstate boundary conditions at the origin and at $\infty$, irrespective of whether $l_1$ is an integer or not,  but it does not correspond to a physical state unless the angular momentum is an integer. However (\ref{eq:I2}) and (\ref{eq:I6}) show that the radial equation for $l_1=2l_2+\frac{1}{2},\ \nu_1=2$ is related to the radial equation for $\nu_2=-1$ and integer $l_2$. This gives rise to the identification that the radial equation for the oscillator potential for unphysical angular momenta $l_1=\frac{1}{2},\ \frac{5}{2}, \frac{9}{2},..$ at energies $\epsilon_1=4(n+l_2+1)$ can be transformed to a new radial equation using the coordinate transformation $\rho_1^2 = \frac{\rho_2}{n+l_2+1}$, leading to (\ref{eq:I5}) with $\nu_2=-1$, which has the solutions 
\begin{equation}
 u_2\ = \rho_2^{l_2+1}\ \Big(\exp \frac{-\rho_2}{2(n+ l_2+1)}\Big)\ M\Big(-n,2l_2+2, \frac{\rho_2}{n+l_2+1}\Big)
\end{equation}
which correspond to true bounstate solutions in an attractive Coulomb potential for physical values of angular momenta $l_2=0,1,2,...$ with boundstate energies $ \epsilon_2 = -\frac{4}{\epsilon_1^2}= -\frac{1}{4}\ \frac{1}{(n+l_2+1)^2}$. Using eqs.(\ref{eq:I0}),(\ref{eq:I1}) and (\ref{eq:I4})-(\ref{eq:I6}) the relationship between the solutions $u_1$ and $u_2$ can be established to be
\begin{equation}
u_2(\rho_2)\ \sim \rho_2^{\frac{1}{4}}\ u_1(\rho_1),\ \ \rho_2\ =\ \frac{\epsilon_1}{4} \rho_1^2,\ \ l_1\ =\ 2 l_2 + \frac{1}{2}
\end{equation}
This relationship is the well known SHO-Coulomb correspondence  which establishes a relation between the oscillator solutions expressed in terms of polynomials and the Coulomb boundstate solutions expressed in terms of  polynomials. It is to be emphasized that unlike the examples given earlier, the oscillator-Coulomb correspondence is not a relation between two physical systems in Quantum Mechanics but is a relation between two solutions of two different radial equations. However in Classical Mechanics the restriction of the angular momentum to values which are integer units of $\hbar$ does not apply and there is an exact equivalence of the SHO as a physical system to the negative energy states of a physical system corresponding to bounded motion in a Coulomb potential. The exact correspondence of the classical orbits of the SHO and the bound orbits in a Coulomb potential will be further examined in section 4. 

The potential $V_1=\rho_1^{\nu_1}$ with positive values of the exponent is a confining potential which has only bound states and no scattering states. The potential $V_2=-\rho_2^{\nu_2}$ with a negative exponent is a singular potential which has both bound states and scattering states. The transformation we have studied in this section establishes a correspondence between the bound states of $V_1$ and $V_2$ when the exponents and the angular momenta fulfill certain conditions. However the significance of the transformation of the scattering states of $V_2$ when there is no corresponding scattering states of $V_1$ remains to be understood. 

\section{Semi-classical analysis of power law potentials}

The correspondence between the radial Schr{\"{o}}dinger equations for two related power law potentials suggests that there must be a similar correspondence between the classical action integrals too which would induce a relation between the WKB quantization conditions for the two power law potentials. In semi-classical analysis the centrifugal potential in Quantum Mechanics is replaced using the Langer modification $l(l+1)\ \rightarrow\ (l+\frac{1}{2})^2$ is implemented. In Classical Mechanics the centrifugal potential is $\frac{l^2}{\rho^2}$. It is interesting to note that the Langer modification of the angular momentum of the classical value of the angular momentum $l$ through the replacement $l\rightarrow (l+\frac{1}{2})$ in the semi-classical analysis naturally arises in the study of the transformations linking the radial Schr{\"{o}}dinger equations for two power law potentials. The action integral associated with the attractive singular potential $v_2=-\rho_2^{\nu_2}$, $-2\le \nu_2 \le 0$, for the negative energy $\epsilon_2$ and angular momentum $l_2$ is
\begin{equation}
S\ = \int_{t_{1,2}}^{t_{2,2}} {\sqrt {\epsilon_2\ +\ \rho_2^{\nu_2}\ -\ \frac{(2l_2+1)^2}{4\rho_2^2}}}\ \ d\rho_2 \label{eq:I12}
\end{equation}
where $t_{1,2}$ and $t_{2,2}$ are the classical turning points where the integrand vanishes. Under the transformation to a new variable
\begin{equation}
\rho_2\ =\ {\sqrt {-\frac{1}{\epsilon_2}}}\ \Big(\frac{2+\nu_2}{2}\Big)\ \rho_1^{\frac{2}{2+\nu_2}}\ \ \ ,\ \ \ d\rho_2\ =\ {\sqrt {-\frac{1}{\epsilon_2}}}\ \rho_1^{\frac{-\nu_2}{2+\nu_2}}\ d\rho_1 \label{eq:I13}
\end{equation}
the action integral transforms to
\begin{equation}
S\ = \int_{t_{1,1}}^{t_{2,1}} {\sqrt{-\rho_1^{\frac{-2\nu_2}{2+\nu_2}} + \Big(\frac{2+\nu_2}{2}\Big)^{\nu_2} \Big(-\frac{1}{\epsilon_2}\Big)^{\frac{2+\nu_2}{2}}\ -\ \frac{(2l_2+1)^2}{4\rho_1^2} \Big(\frac{2}{2+\nu_2}\Big)^2 }} \ d\rho_1
\end{equation}
where $t_{1,1}$ and $t_{2,1}$ are the turning points in the transformed coordinate $\rho_1$. If we now identify a new exponent, a new energy and a new angular momentum through the definitions
\begin{equation}
\nu_1 = -\frac{2\nu_2}{2+\nu_2}\ ,\ \epsilon_1 = \Big(-\frac{1}{\epsilon_2}\Big)^{\frac{2+\nu_2}{2}}\ \Big(\frac{2+\nu_2}{2}\Big)^{\nu_2}\ \ ,\ 2 l_1 + 1 = \Big(\frac{2}{2+\nu_2}\Big) (2 l_2 + 1) \label{eq:I14}
\end{equation}
which are exactly the same transformations as identified in eqs. (\ref{eq:I2}),(\ref{eq:I4}) and (\ref{eq:I6}) from the study of the radial Schr{\"{o}}dinger equation, then $S$ can be brought to the form
\begin{equation}
S\ = \int_{t_{1,1}}^{t_{2,1}} {\sqrt {\epsilon_1\ -\ \rho_1^{\nu_1}\ -\ \frac{(2l_1+1)^2}{4\rho_1^2}}}\ \ d\rho_1 \label{eq:I15}
\end{equation}
which can now be interpreted as the action integral associated with the potential $v_1=\rho_1^{\nu_1}$ for the energy $\epsilon_1$ and angular momentum $l_1$ and $t_{1,1}$ and $t_{2,1}$ are the  classical turning points of the new potential.  It is also possible to start from (\ref{eq:I15}) and use the inverse transformation
\begin{align}
\rho_1\ &=\ {\sqrt {\frac{1}{\epsilon_1}}}\ \Big(\frac{2+\nu_1}{2}\Big)\ \rho_2^{\frac{2}{2+\nu_1}}\ \ \ ,\ \ \ d\rho_1\ =\ {\sqrt {\frac{1}{\epsilon_1}}}\ \rho_2^{\frac{-\nu_1}{2+\nu_1}}\ d\rho_2 \label{eq:J13}\\
\nu_2 &= -\frac{2\nu_1}{2+\nu_1}\ ,\ \epsilon_2 = - \Big(\frac{1}{\epsilon_1}\Big)^{\frac{2+\nu_1}{2}}\ \Big(\frac{2+\nu_1}{2}\Big)^{\nu_1}\ \ ,\ 2 l_2 + 1 = \Big(\frac{2}{2+\nu_1}\Big) (2 l_1 + 1) \label{eq:J14}
\end{align}
and recover (\ref{eq:I12}). The exact identity of the action integrals for the two related power law potentials under the mapping defined by (\ref{eq:I13}) and (\ref{eq:I14}) or (\ref{eq:J13}) and (\ref{eq:J14}) then leads to a mapping of the eigenvalues through the WKB quantization condition appropriate to the radial equation. For non-singular potentials the quantization condition can be shown to be $S= (n-\frac{1}{4}) \pi,\ n=1,2,3,..$\ . However the quantization condition for singular potentials requires subtle handling. The action integral for singular potentials  given by (\ref{eq:I12}) can be performed when $l_2=0$. 
The semi-classical quantization condition for singular potentials and positive integer values of $l_2$ has been considered carefully by Quigg and Rosner (p203) and using their analysis it can be  shown that the semi-classical spectrum of the singular potential $v_2$ is given by
\begin{equation}
\epsilon_2(n,l_2)\ =\ -\Big[A_2\ \Big(n - \frac{1}{2}\   \frac{1+\nu_2-2l_2}{2+\nu_2}\Big)\Big]^{\frac{2\nu_2}{2+\nu_2}}\ \ ,\ \ A_2\ =\ 2\ {\sqrt{\pi}}\ \frac{\Gamma\Big(-\frac{1}{\nu_2}\Big)}{\Gamma\Big(-\frac{1}{2}-\frac{1}{\nu_2}\Big)} 
\end{equation}
The equality of the action integrals in (\ref{eq:I12}) and (\ref{eq:I15}) under the mapping given in (\ref{eq:I14}) enables the determination of the WKB eigenvalues of $v_1$ also. The WKB eigenvalues for the potential $v_1$ for integer values of $l_1$ can be given in the form (Quigg and Rosner p205)
\begin{equation}
\epsilon_1(n,l_1)\ =\ \Big[A_1\ \Big(n + \frac{l_1}{2} - \frac{1}{4}\Big)\Big]^{\frac{2\nu_1}{2+\nu_1}} \ \ ,\ \ A_1\ =\ {\sqrt{\pi}}\ (2+\nu_1)\ \frac{\Gamma\Big(\frac{1}{2} + \frac{1}{\nu_1}\Big)}{\Gamma\Big(\frac{1}{\nu_1}\Big)} \label{eq:G9}
\end{equation}
It is simple to to verify that  when the conditions (\ref{eq:G8}) are satisfied then the semi-classical eigenvalues for the potentials $v_1$ and $v_2$ given above satisfy the same condition (\ref{eq:I8}) as that satisfied by the exact eigenvalues from Quantum Mechanics.

\subsection{The approximate degeneracy in the spectra of $v_1=\rho_1^{\nu_1}, \nu_1>0$} 

The WKB spectrum of $v_1$ implies that the spectrum of a confining power law potential $v_1$ depends only on the combination $(2n+l_1)$ and exhibits the same degeneracy as the 3-d oscillator. This WKB result implies that for fixed $l, n>>l$, the exact Quantum Mechanical results for the spectra of power law potentials with positive exponents should also exhibit the same degeneracy. This feature can be understood from the point of view of Super Symmetric Quantum Mechanics (SUSY) which establishes a relation between a deep singular potential and a phase equivalent shallow potential which is constructed by removing a certain number $N$ of the lowest  boundstates of the deep potential so that the spectrum of the shallow potential constructed by this procedure is identical to that of the deep potential except for missing the lowest $N$ eigenvalues of $V_{deep}$. It has been shown that (Baye 1987, Sukumar 1985,2005)
\begin{align}
V_{shallow}^{(N)} \ &=\ V_{deep}\ -2 \frac{\partial^2}{\partial\rho^2} \Big(\ln Det M\Big)\ \ ,\ \ M_{jk}\ =\ \int_0^\rho R_j^{(d)} R_k^{(d)} \  d\rho,\ \ j,k = 1,2,..,N \\
Lt_{\rho\rightarrow 0}\  \ V_{shallow}^{(N)} &\sim \frac{(l + 2N) (l + 2N + 1)}{\rho^2}\ \  {\hbox {if}}\ \ Lt_{\rho\rightarrow 0}\ \  V_{deep} \sim \frac{l(l + 1)}{\rho^2} \label{eq:S1}\\
Lt_{\rho\rightarrow \infty}\  \ V_{shallow}^{(N)} &= \ Lt_{\rho\rightarrow \infty}\ \  V_{deep} 
\end{align}
where $R_j^{(d)}$ are the radial eigenfunctions in the deep potential. The enhanced centrifugal barrier of the shallow potential with a $(2N+l)$ dependence seen in (\ref{eq:S1}) is a short range feature valid only for $\rho<<1$. However for power law potentials with positive exponents the centrifugal potential is not significant in the long range as the $\rho^\nu$ part will dominate in the limit  $\rho\rightarrow\infty$. Hence a shallow potential constructed from a power law potential by removing the lowest bound states by the SUSY procedure may, to a good approximation, be viewed as a power law potential plus a centrifugal potential corresponding to an enhanced angular momentum barrier. These results together with (\ref{eq:G9}) imply that the $n^{th}$ semi-classical eigenvalue of the potential $v_1^{(eff)}=(\rho_1^{\nu_1} +$ the centrifugal potential for angular momentum $l=2l_1)$ depends on the quantum number $(n+l_1),\ n=1,2,..,$  and is degenerate with the $n^{th}$ semi-classical eigenvalue of the 'shallow' potential  ${\bar v}_1^{(l_1)}$ constructed by eliminating the lowest lying $l_1$ eigenstates of the 'deep' potential $v_1=\rho_1^{\nu_1}$  whose eigenvalues depend on the quantum number $n=1,2,..\ $. Similarly the $n^{th}$ semi-classical eigenvalue of the potential $v_1^{(eff)}=(\rho_1^{\nu_1} +$ the centrifugal potential for angular momentum $l=2l_1+1)$ depends on the quantum number $(n+l_1+\frac{1}{2}),\ n=1,2,..,$, and is degenerate with the $n^{th}$ semi-classical eigenvalue of the 'shallow' potential ${\bar v}_1^{(l_1)}$ constructed by eliminating the lowest lying $l_1$ eigenstates of the 'deep' potential $\rho_1^{\nu_1} +\frac{2}{\rho_1^2}$ whose eigenvalues depend on the quantum number $ (n+\frac{1}{2}),\ n=1,2,..\ $. The potentials identified above as having degenerate spectra are not identical but the semi-classical quantization formulae suggest that the difference $\Delta v$ between the potentials with the degenerate spectra does not play a significant role when determining the eigenvalues of states with $n>>l$. Thus the SUSY construction elucidates the feature that in the semi-classical limit, $l>>1, n>>l,$ the eigenvalues of the power law potentials with positive exponents will exhibit a $(2n+l)$ dependence. 

This result is also in agreement with the exactly solvable problem of the limit $\nu_1\rightarrow \infty$ which corresponds to a particle confined to move inside a unit sphere, ({\it {i.e}}) a vanishing potential inside a sphere radius $\rho_1=1$ and an infinite potential  at $\rho_1=1$ representing an impenetrable wall. The radial equation for a particle inside a spherical box  may be solved and the radial solution which is regular at the origin is
\begin{equation}
R\ =\ {\sqrt{\rho_1}}\ J_{l+\frac{1}{2}}({\sqrt{\epsilon_1}}\rho_1)\ \ ,\ \  Lt_{k\rho\rightarrow \infty} \ J_{l+\frac{1}{2}}(k\rho) \sim \sin\Big(k\rho - l \frac{\pi}{2}\Big)
\end{equation}
The eigenvalues arise from the requirement that the radial solution should vanish at the infinite wall, ({\it{i.e}}) when $\rho_1\rightarrow 1$. Hence the eigenvalues are related to the zeros of Spherical Bessel functions. When the argument of the Bessel function is large the asymptotic limit of the zeros can be estimated using McMahon's expansion (Abramowitz and Stegun p371) which locates the $n^{th}$ zero $Z_n$ at
\begin{equation}
Z_n\  \sim \ \beta\ -\ \frac{\sigma-1}{8\beta}\ -\  O\Big(\beta^{-3}\Big)\ \ ,\ \ \beta\ =\ \Big(n +\frac{l}{2}\Big) \pi\ \ , \ \ \sigma\ =\ (2l+1)^2
\end{equation}
The semi-classical energy eigenvalues for a particle in a spherical box are given by 
\begin{equation}
Lt_{n\rightarrow\infty}\ \epsilon_1(n,l)\ \sim\ \Big((2n\ +\ l)\ \frac{\pi}{2}\Big)^2
\end{equation}
which clearly exhibits a (2n+l) degeneracy. 

The (2n+l) degeneracy is an exact result for the 3-d oscillator but it is also a very good approximation for all power law potentials with positive exponents in the semi-classical limit of large quantum numbers $n >> l >> 1$. Gaussian potentials are often used to describe short range potentials in Nuclear Physics. Numerical calculation of the eigenvalues of a Gaussian potential shows that the eigenvalue spectrum exhibits a (2n+l) degeneracy to a good approximation. It is therefore possible to conjecture, without explicit proof, that the (2n+l) degeneracy is a good approximation for a large class of potentials in the limit of large quantum numbers.

\section{Classical Mechanics of power law potentials}

In classical Mechanics the energy and angular momentum conservation laws lead to an equation for the orbit. For a particle in a potential $V_1=\rho_1^{\nu_1}$, $0\le\nu_1\le\infty$, with angular momentum $l_1$ and positive energy $\epsilon_1$,
\begin{align}
\epsilon_1\ &=\ \Big(\frac{\partial\rho_1}{\partial t}\Big)^2\ +\ \frac{l_1^2}{\rho_1^2}\ +\ \rho_1^{\nu_1}\ ,\ \ \Big(\frac{\partial\rho_1}{\partial t}\Big)^2\ \frac{\partial\theta_1}{\partial t}\ =\ l_1\ =\ {\hbox {constant}} \\
\theta_1\ &=\ l_1\ \int \frac{d\rho_1}{\rho_1^2}\ \Big(\epsilon_1 - \frac{l_1^2}{\rho_1^2} - \rho_1^{\nu_1}\Big)^{-\frac{1}{2}} \label{eq:I16}
\end{align}
Similarly for a particle in an attractive singular potential $V_2=-\rho_2^{\nu_2}$, $-2\le\nu_2\le0$, with angular momentum $l_2$ and negative energy $\epsilon_2$ moving in a bound orbit
\begin{align}
-|\epsilon_2|\ &=\ \Big(\frac{\partial\rho_2}{\partial t}\Big)^2\ +\ \frac{l_2^2}{\rho_2^2}\ -\ \rho_2^{\nu_2}\ ,\ \ \Big(\frac{\partial\rho_2}{\partial t}\Big)^2\ \frac{\partial\theta_2}{\partial t}\ =\ l_2\ =\ {\hbox {constant}} \\
\theta_2\ &=\ l_2\ \int \frac{d\rho_2}{\rho_2^2}\ \Big(-|\epsilon_2| - \frac{l_2^2}{\rho_2^2} + \rho_2^{\nu_2}\Big)^{-\frac{1}{2}} \label{eq:I17}
\end{align}
It is to be noted that in Classical Mechanics the centrifugal term is $\frac{l^2}{{\rho}^2}$ without the Langer modification used in the Semi-classical analysis. The equation of the orbit can also be found directly from the classical action integral by differentiation 
\begin{align}
\theta_1\ &=\ -\frac{\partial S_1}{\partial l_1}\ \ ,\ S_1\ =\ \int_{t_{1,1}}^{t_{2,1}} {\sqrt {\epsilon_1\ -\ \rho_1^{\nu_1}\ -\ \frac{l_1^2}{\rho_1^2}}}\ \ d\rho_1 \\
\theta_2\ &=\ -\frac{\partial S_2}{\partial l_2}\ \ ,\ S_2\ =\  \int_{t_{1,2}}^{t_{2,2}} {\sqrt {\epsilon_2\ +\ \rho_2^{\nu_2}\ -\ \frac{l_2^2}{\rho_2^2}}}\ \ d\rho_2
\end{align}
where $(t_{1,1},t_{1,2})$ and $(t_{2,1},t_{2,2})$ are the classical turning points in the two potentials. It may be shown by starting from the expression for $S_2$ and using the transformation of variables in (\ref{eq:I13}) and the mapping
\begin{equation}
\nu_1 = -\frac{2\nu_2}{2+\nu_2}\ ,\ \epsilon_1 = \Big(-\frac{1}{\epsilon_2}\Big)^{\frac{2+\nu_2}{2}}\ \Big(\frac{2+\nu_2}{2}\Big)^{\nu_2}\ \ ,\  l_1  = \Big(\frac{2}{2+\nu_2}\Big)  l_2 
\end{equation}
that $S_2=S_1$. The mapping given above differs slightly from the transformation in (\ref{eq:I14}) due to the fact that in the absence of Langer modification the relation between the angular momenta becomes $\nu_1 l_2= -\nu_2 l_1$. The equality of the classical actions in the two related power law potentials induces the relation
\begin{equation}
\theta_2\ =\ -\frac{\partial S_2}{\partial l_2}\ =\ -\frac{\partial l_1}{\partial l_2} \frac{\partial S_1}{\partial l_1}\ =\ -\frac{\nu_1}{\nu_2}\ \theta_1
\end{equation}
This relation may also be established by staring from (\ref{eq:I17}) and using a transformation of coordinates to relate it to (\ref{eq:I16}). 

The general correspondence between the orbital equations in the two related potentials may be given in the form
\begin{align}
\rho_1\ &=\ F(\theta_1,\ l_1,\ \epsilon_1) \ ,\ \rho_2\ =\ \frac{2}{2+\nu_1} \frac{1}{{\sqrt{-\epsilon_2}}}\ \rho_1^{\frac{2+\nu_1}{2}}\\
\theta_1\ &=\ \frac{2\theta_2}{2+\nu_1}\ ,\ l_1 = \frac{2+\nu_1}{2} \ l_2\ ,\ \epsilon_1 =\ \Big(\frac{2+\nu_1}{2}\Big)^{\frac{2\nu_1}{2+\nu_1}}\ \Big(-\epsilon_2\Big)^{-\frac{2}{2+\nu_1}} 
\end{align}
where the orbital function $F(\theta_1,l_1,\epsilon_1)$ is found by integration of (\ref{eq:I16}). The mapping between the parameters of the two orbits has a symmetrical structure and so it is also possible to start from the orbital equation for $V_2$ and find the equivalent orbit of the transformed potential  $V_1$ using  the mapping
\begin{align}
\rho_2\ &=\ F(\theta_2,\ l_2,\ \epsilon_2) \ ,\ \rho_1\ =\ \frac{2}{2+\nu_2} \frac{1}{{\sqrt{\epsilon_1}}}\ \rho_2^{\frac{2+\nu_2}{2}}\\
\theta_2\ &=\ \frac{2\theta_1}{2+\nu_2}\ ,\ l_2 = \frac{2+\nu_2}{2} \ l_1\ ,\ \epsilon_2 =\ -\Big(\frac{2+\nu_2}{2}\Big)^{\frac{2\nu_2}{2+\nu_2}}\ \Big(\epsilon_1\Big)^{-\frac{2}{2+\nu_2}} 
\end{align}

As an example of the orbital relation we consider the orbit in a three dimensional oscillator potential $V_1=\rho_1^2$ for which the equation of the orbit may be found by  direct integration of (\ref{eq:I16}) and may be shown to be of the form
\begin{equation}
\frac{1}{\rho_1^2}\ =\ \frac{\epsilon_1}{2 l_1^2}\ \Big(1\ -\ {\sqrt{1 - \frac{4 l_1^2}{\epsilon_1^2}}}\ \cos 2\theta_1\Big) \label{eq:I18}
\end{equation}
Only bound orbits exist with $\epsilon_1\ge0$. Under the transformation
\begin{equation}
\rho_2 = \frac{1}{2{\sqrt{-\epsilon_2}}}\ \rho_1^2\ ,\ \ \theta_1 = \frac{\theta_2}{2}\ ,\ l_1 = 2l_2\ ,\ \epsilon_1 = 2 (-\epsilon_2)^{-\frac{1}{2}} \label{eq:J18}
\end{equation}
we obtain 
\begin{equation}
\frac{1}{\rho_2}\ =\ \frac{1}{2 l_2^2}\ \Big(1\ -\ {\sqrt{1 + 4\epsilon_2 l_2^2}}\ \cos\theta_2\Big) \label{eq:I19}
\end{equation}
which is a bound orbit in the Coulomb potential $V_2=-\rho_2^{-1}$ with $\epsilon_2\le0$. The inverse transformation in this case is
\begin{equation}
\rho_1 = \frac{2}{{\sqrt{\epsilon_1}}}\ \rho_2^{\frac{1}{2}}\ ,\ 
\theta_2\ =\ 2\theta_1\ ,\ l_2 = \frac{l_1}{2}\ ,\ \epsilon_2 =\ -\Big(\frac{1}{2}\Big)^{-2}\ \Big(\epsilon_1\Big)^{-2} \label{eq:J19}
\end{equation}
which transforms the orbit in (\ref{eq:I19}) back to the orbit given in (\ref{eq:I18}).

The orbit in an attractive Coulomb potential given in (\ref{eq:I19}) is also the correct solution to (\ref{eq:I17}) for positive values of $\epsilon_2$ and may be identified as a hyperbola. The mapping given in (\ref{eq:J18}) indicates that a positive value for $\epsilon_2$ corresponds to imaginary values of $\epsilon_1$ and the solution for the oscillator given in (\ref{eq:I18}) becomes complex and this complex solution for the orbit would require an interpretation. This difficulty is perhaps also related to the question raised at the end of section 2.1 regarding the transformation of the radial Schr{\"{o}}dinger equation for positive energy scattering states of singular potentials $V_2$ to the radial Schr{\"{o}}dinger equation for imaginary $\epsilon_1$  states of confining potentials $V_1$.

\section{References}

[1] C.Quigg and J.L.Rosner, 1979 {\it{Physics Reports}} {\bf 56}  167-235.

[2] M.Abramowitz and I.Stegun, 1965 {\it{Handbook of Mathematical Functions}} Dover: New York 504

[3] C.V.Sukumar, 1985 {\it{J.Phys. A: Math. Gen.}} {\bf 18} 2937-56

[4] D.Baye, 1987 {\it{Physical Review Letters}} {\bf 58} 2738-41

[5] C.V.Sukumar, 2005 {\it{Latin American School of Physics XXXV, ELAF, Supersymmetries in Physics and its applications, edited bt R.Bijker, et al.}} AIP conference proceedings {\bf 744} 166-235

\end{document}